\documentstyle[11pt]{article}

\begin{document}

\title{\bf Universal regularization prescription for Lovelock AdS gravity}
\author{Georgios Kofinas$^a$ and Rodrigo Olea$^b$ \medskip \\
{\small {\em $^a$ Department of Physics and Institute of Plasma
Physics, University of Crete, 71003 Heraklion, Greece,}} \\
{\small {\em $^{b}$ INFN, Sezione di Milano, Via Celoria 16,
I-20133, Milano, Italy.}}\\
{\small {\tt gkofin@phys.uoa.gr, rodrigo.olea@mi.infn.it}}}

\maketitle


\begin{abstract}
 A definite form for the boundary term that produces
the finiteness of both the conserved quantities and Euclidean action
for any Lovelock gravity with AdS asymptotics is presented. This
prescription merely tells even from odd bulk dimensions, regardless
the particular theory considered, what is valid even for
Einstein-Hilbert and Einstein-Gauss-Bonnet AdS gravity. The boundary
term is a given polynomial of the boundary extrinsic and intrinsic
curvatures (also referred to as Kounterterms series). Only the
coupling constant of the boundary term changes accordingly, such
that it always preserves a well-posed variational principle for
boundary conditions suitable for asymptotically AdS spaces. The
background-independent conserved charges associated to asymptotic
symmetries are found. In odd bulk dimensions, this regularization
produces a generalized formula for the vacuum energy in Lovelock AdS
gravity. The standard entropy for asymptotically AdS black holes is
recovered directly from the regularization of the Euclidean action,
and not only from the first law of thermodynamics associated to the
conserved quantities.

\end{abstract}


\section{Introduction}

It is believed that the Einstein-Hilbert action is just the first
term in the derivative expansion in a low energy effective theory.
In general, higher order quantum corrections to gravity might
appear, whose corresponding couplings are unknown until now. Among
the higher derivative gravity theories, Lovelock gravity
\cite{lovelock} possesses some special features: it leads to field
equations which are up to and linear in second derivatives of the
metric, it obeys generalized Bianchi identities which ensure energy
conservation, and it is known to be free of ghosts when expanded on
a flat space, avoiding problems with unitarity \cite{ghost-free}.

In presence of cosmological constant, the Euclidean continuation of
the bulk gravity action and the conserved quantities are in general
divergent. In the AdS/CFT context \cite{maldacena}, one deals with
the regularization problem for Einstein-Hilbert action by adding
local functionals of the boundary metric (Dirichlet counterterms)
\cite{bala}. Because of this dependence, they preserve a
well-defined variational action principle for a Dirichlet boundary
condition on the metric (achieved through the Gibbons-Hawking term)
when varied. However, the systematic construction \cite{henniskende}
that provides the form of the counterterms becomes cumbersome for
high enough dimensions, what has prevented from finding a general
pattern for the series for any dimension until now. In Lovelock
gravity, it is expected that the holographic renormalization
procedure would be even more complicated.

An alternative regularization scheme has been proposed for
Einstein-Hilbert \cite{OleaJHEP, OleaKounter}, Einstein-Gauss-Bonnet
\cite{Kofinas-Olea}, and Chern-Simons
\cite{Mora-Olea-Troncoso-Zanelli-CS} gravity theories with AdS
asymptotics. It considers the addition to the bulk action of
boundary terms with dependence on the extrinsic curvature. In this
paper, we show that this prescription is universal for all
Lovelock-AdS theories, attaining a regularized action and finiteness
of the conserved charges .

\section{Lovelock gravity}

In $D=d+1$ dimensions, the Lovelock action reads
\begin{equation}
\!\!\!\!\!\!\!\!\!\!\!\!\!\!\!I_{D}=\frac{1}{16\pi G_{\!D}}\int_{M}
\!\sum_{p=0}^{[(\!D-1\!)\!/2]}\alpha _{p}L_{p}+c_{d}\int_{\partial
M}\!\!\!\!B_{d}, \label{lovelock}
\end{equation}
where $L_{p}$ corresponds to the dimensional continuation of the
Euler term in $2p$ dimensions
\begin{eqnarray}
\!\!\!\!\!\!\!\!\!\!\!\!\!\!\!L_{p} &=&\frac{1}{(D-2p)!}\epsilon
_{A_{1}...A_{D}}\hat{{\mathcal{R}}}^{A_{1}A_{2}}...\hat{{\mathcal{R}}}
^{A_{2p-1}A_{2p}}e^{A_{2p+1}}...e^{A_{D}},  \label{Lp} \\
&=&\frac{1}{2^{p}}\sqrt{-\mathcal{G}}\,\delta _{\left[ \mu
_{1}\cdots \mu _{2p}\right] }^{\left[ \nu _{1}\cdots \nu
_{2p}\right] }\, \hat{R}_{\nu _{1}\nu _{2}}^{\mu _{1}\mu _{2}}\cdots
\hat{R}_{\nu _{2p-1}\nu _{2p}}^{\mu _{2p-1}\mu _{2p}}\,d^{D}x\,.
\label{Lptensor}
\end{eqnarray}
Hatted curvatures stand for $D$-dimensional ones. The orthonormal
vielbein $ e^{A}\!=\!e_{\,\,\,\mu }^{A}dx^{\mu }$ produces the
spacetime metric by ${\mathcal{G}}_{\mu\nu}\!=\!\eta
_{AB}\,e_{\,\,\,\mu }^{A}e_{\,\,\,\nu }^{B}$ and the curvature
2-form is defined as $\hat{{\mathcal{R}}}^{AB}\!=\!d\omega
^{AB}\!+\!\omega _{\,\,\,\,C}^{A}\omega ^{CB}$ in terms of the spin
connection one-form $ \omega ^{AB}\!=\!\omega _{\,\,\,\,\,\,\,\,\mu
}^{AB}dx^{\mu }$ and related to the spacetime Riemman tensor by
$\hat{{\mathcal{R}}}^{AB}\!=\!\frac{1}{2}\hat{R}_{\mu \nu }^{\kappa
\lambda }e_{\,\,\,\kappa}^{A}e_{\,\,\,\lambda}^{B}dx^{\mu }dx^{\nu
}$. Wedge products are omitted throughout. The first term in the
Lovelock series corresponds to the cosmological term $
L_{0}\!=\!\sqrt{-\mathcal{G}}\,\,d^{D}x$,
$L_{1}\!=\!\sqrt{-\mathcal{G}}\,\hat{R}\,d^{D}x$ is the
Einstein-Hilbert term, $L_{2}\!=\!
\sqrt{-\mathcal{G}}\,(\hat{R}_{\mu \nu \kappa \lambda }\hat{R}^{\mu
\nu \kappa \lambda }\!-\!4\hat{R}_{\mu \nu }\hat{R}^{\mu \nu
}\!+\!\hat{R}^{2})\,d^{D}x$ is the Gauss-Bonnet term, etc. The first
coefficients are $\alpha
_{0}\!=\!-2\Lambda\!=\!\frac{(D-1)(D-2)}{\ell^{2}}$, $\alpha
_{1}\!=\!1$, whereas all the other $\alpha _{p}$'s are arbitrary.

The action (\ref{lovelock}) appears supplemented by a boundary term
$B_{d}$. We shall display below the universal form of $B_{d}$ for
any Lovelock theory with AdS asymptotia that regularizes both the
conserved quantities and the Euclidean action.

The equation of motion for a generic Lovelock gravity (with zero
torsion) is obtained varying with respect to the metric and takes
the form
\begin{equation}
E_{\mu }^{\nu }=\sum_{p=0}^{[(\!D-1\!)\!/2]}\frac{\alpha
_{p}}{2^{p}} \,\,\delta _{\left[ \mu \mu _{1}\cdots \mu _{2p}\right]
}^{\left[ \nu \nu _{1}\cdots \nu _{2p}\right] }\,\hat{R}_{\nu
_{1}\nu _{2}}^{\mu _{1}\mu _{2}}\cdots \hat{R}_{\nu _{2p-1}\nu
_{2p}}^{\mu _{2p-1}\mu _{2p}}=0. \label{EOMlovelock}
\end{equation}

The vacua of a given Lovelock theory are defined as the maximally
symmetric spacetimes that are globally of constant curvature. We
will assume that all the corresponding cosmological constants are
real and negative, i.e.,
\begin{equation}
\Lambda _{e\!f\!\!f}=-\frac{(D\!-\!1)(D\!-\!2)}{2\ell
_{\!e\!f\!\!f}^{2}}, \label{Lambdaeff}
\end{equation}
where $\ell _{\!e\!f\!\!f}$ is defined as the effective AdS radius
given by the solutions to the equation
\begin{equation}
\sum\limits_{p=0}^{[(\!D-1\!)\!/2]}\!\,\frac{\alpha
_{p}}{(D\!-\!2p\!-\!1)!}(-\ell _{\!e\!f\!\!f}^{-2})^{p}=0.
\label{elleff}
\end{equation}

In the present paper, we will consider spacetimes whose asymptotic
behavior tends to the one of a locally AdS space, described in terms
of its curvature by the condition
\begin{equation}
\hat{R}_{\mu \nu }^{\kappa \lambda}+\frac{1}{\ell
_{\!e\!f\!\!f}^{2}}\delta _{\lbrack \mu \nu ]}^{[\kappa \lambda ]}=0
\label{AAdSR}
\end{equation}
at the boundary $\partial M$, or equivalently, $\hat{{\mathcal{R}}}
^{AB}\!+\!(e^{A}e^{B})/\ell _{\!e\!f\!\!f}^{2}=0$ in differential
forms language. It is important to stress that this is a generic
(local) condition that does not fix completely the form of the
metric.

In principle, it is not clear whether the holographic
renormalization procedure might provide a systematic algorithm to
regularize a generic Lovelock-AdS theory, because of the increasing
complexity of the field equations respect to the Einstein-Hilbert
case. The alternative construction in this paper represents a way of
circumventing the difficulties of the standard method because, as we
shall see below, it does not make use of the full expansion of the
asymptotic metric. Indeed, we will only consider the leading order
for the fields induced by this expansion to identify suitable
boundary conditions for the variational problem in AAdS gravity.

Without loss of generality, we write down the line element in Gauss-normal
coordinates
\begin{equation}
ds^{2}=N^{2}(\rho )d\rho ^{2}+h_{ij}(x,\rho )dx^{i}dx^{j},  \label{radial}
\end{equation}
that can be obtained from a generic radial ADM foliation by
gauge-fixing the shift functions $N^{i}=0$. A definite choice of the
lapse and the boundary metric generically describes AAdS spaces in
Lovelock gravity. Indeed, taking the lapse and the boundary metric
as
\begin{eqnarray}
N\! &=&\!\ell _{\!e\!f\!\!f}/2\rho ,  \label{NFG} \\
h_{ij}\! &=&\!g_{ij}(x,\rho )/\rho ,  \label{hFG}
\end{eqnarray}
where $g_{ij}(x,\rho )$ accepts a regular Fefferman-Graham expansion
\cite {fg}
\begin{equation}
g_{ij}\!(x,\rho )=g_{(0)ij}\!(x)+\rho\,g_{(1)ij}\!(x)+\rho
^{2}\,g_{(2)ij}\!(x)+...  \label{gFG}
\end{equation}
identically satisfies the condition (\ref{AAdSR}) at the conformal boundary $%
\rho \!=\!0$. Here, $g_{(0)}$ is the boundary data of an
initial-value problem, governed by the equations of motion written
in the frame (\ref{radial})-(\ref{hFG}). However, even for
Einstein-Hilbert theory, solving the coefficients $g_{(k)}$ in
series (\ref{gFG}) as covariant functionals of $g_{(0)}$ is only
possible for low enough dimensions. Moreover, for theories where
eq.(\ref{elleff}) has a single root, the equations of motion posses
a multiple zero in a unique AdS vacuum \cite{DCBH, BHScan}. This
causes the first nontrivial relation for a given coefficient
$g_{(k)}$ to appear at a higher order in $\rho $, what substantially
increases the complexity of the equations. Therefore, one can expect
that the extreme nonlinearity of the field equations in Lovelock-AdS
gravity would turn impractical the application of holographic
renormalization method to this class of theories.

In what follows, we propose a universal form of the boundary terms
that make both the conserved charges and the Euclidean action finite
in Lovelock-AdS gravity. This construction does not make use of the
full Fefferman-Graham form of the metric (\ref{radial})-(\ref{gFG})
for AAdS spacetimes, but simply considers the leading-order terms in
the expansion of the relevant fields.

\section{$D=2n+1$ dimensions}

In Einstein-Hilbert-AdS gravity, the standard regularization using
Dirichlet counterterms reveals some differences between odd and
even-dimensional cases. Indeed, it is only in odd (bulk) dimensions
that a vacuum energy for AdS spacetime appears. The quasilocal
stress tensor derived from the regularized action features a trace
anomaly only in odd dimensions, as well, what can be traced back to
a logarithmic contribution in the FG expansion (\ref{gFG}).

In the alternative regularization known as Kounterterms method, the
existence of a vacuum energy for Einstein-Hilbert \cite{OleaKounter}
and Einstein-Gauss-Bonnet \cite{Kofinas-Olea} AdS gravity in odd
dimensions is a consequence of a different form of the boundary
terms respect the even-dimensional case. Ultimately, the difference
in the prescription for the regularizing boundary terms is linked to
the existence of topological invariants of the Euler class whose
construction is only possible in even dimensions \cite{four, even}.

The standard Dirichlet counterterms consider the addition to the
action of local, covariant functional of the boundary metric
$h_{ij}$ and the intrinsic curvature $R_{ij}^{kl}(h)$. In the
present formulation, the boundary term $B_{d}$ in
eq.(\ref{lovelock}) will depend also on the extrinsic curvature
$K_{ij}$, defined in the frame (\ref{radial}) by
\begin{equation}
K_{ij}=-\frac{1}{2N}\partial _{\rho }h_{ij}\,,  \label{KdefGC}
\end{equation}
and, because of this dependence, we will refer to it as $\emph{Kounterterms}$
series.

The explicit form the Kounterterms $B_{2n}$ adopt in any
odd-dimensional Lovelock-AdS gravity can be written in a compact way
as
\begin{eqnarray}
B_{\!2n}=2n\sqrt{\!-h}\!\int_{0}^{1}\!\!dt\!\!\int_{0}^{t}\!\!
\!\!ds\,\delta _{\lbrack \!j_{\!1}...
j_{\!2\!n\!-\!1}\!]}^{[i_{\!1}\!...\!i_{\!2\!n\!-\!1}\!]}
K_{i_{\!1}}^{j_{\!1}}\!\Big(\!{\frac{1}{2}}\!R_{i_{2}i_{3}}^{j_{2}j_{3}}\!
-t^{2}\!K_{i_{2}}^{j_{2}}\!K_{i_{3}}^{j_{3}}\!+\!\frac{s^{2}}{\ell
_{\!e\!f\!\!f}^{2}}\delta _{i_{2}}^{j_{2}}\!\delta
_{i_{3}}^{j_{3}}\!\!\Big)\times\cdots \nonumber \\
\cdots\times\Big(\!{\frac{1}{2}}\!R_{i_{2n\!-\!2}i_{2n\!-\!1}}^{j_{2n\!-\!2}j_{2n
\!-\!1}}\!-t^{2}\!K_{i_{2n\!-\!2}}^{j_{2n\!-\!2}}\!K_{i_{2n\!-
\!1}}^{j_{2n\!-\!1}}\!+\!\frac{s^{2}}{\ell _{\!e\!f\!\!f}^{2}}\delta
_{i_{2n\!-\!2}}^{j_{2n\!-\!2}}\delta
_{i_{2n\!-\!1}}^{j_{2n\!-\!1}}\!\!\Big) d^{2n}\!x,
\!\!\label{B2ntensor}
\end{eqnarray}
which, when expanded, produces a polynomial in the intrinsic and
extrinsic curvatures whose relative coefficients are obtained
performing the above parametric integrations
\begin{equation}
B_{2n}=n!\,\sqrt{\!-h}\sum_{p=0}^{n-1}\frac{\left(2n\!-\!2p\!-\!3\right)
!!}{\ell ^{2(n-1-p)}}b_{2n}^{(p)},  \label{B2n-b2n}
\end{equation}
where
\begin{equation}
b_{2n}^{(p)}=\,\delta _{\lbrack j_{1}\cdots j_{2p+1}]}^{[i_{1}\cdots
i_{2p+1}]}\sum_{q=0}^{p}\frac{(-1)^{p-q}\,}{(p\!-\!q)!\,q!}\frac{2^{n-(p+q+1)}}{%
n\!-\!q}\,R_{i_{1}i_{2}}^{j_{1}j_{2}}\cdots
R_{i_{2q-1}i_{2q}}^{j_{2q-1}j_{2q}}\,K_{i_{2q+1}}^{j_{2q+1}}\cdots
K_{i_{2p+1}}^{j_{2p+1}}\,.  \label{b2np}
\end{equation}
The tensorial formula of the boundary terms (\ref{B2ntensor}),
adapted to a radial foliation of the spacetime, can be cast into a
fully Lorentz-covariant $2n-$form with the definition of the second
fundamental form $\theta ^{AB}\!=\!n^{A}K^{B}\!-\!n^{B}K^{A}$,

\begin{eqnarray}
B_{2n} =2n\int_{0}^{1} dt\int_{0}^{t} ds \,\epsilon
_{\!a_{\!1}...a_{2n}}\!K^{a_{1}}e^{a_{2}}\!\Big(\!{\mathcal{R}}
^{a_{3}a_{4}}\!\!-\!t^{2}\!K^{a_{3}}\!K^{a_{4}}\!\!+\!\frac{s^{2}}{\ell
_{\!e\!f\!\!f}^{2}}e^{a_{3}}e^{a_{4}}\!\!\Big)\times \cdots \nonumber\\
\cdots \times \Big(\!{\mathcal{R}}
^{a_{2n-1}a_{2n}}\!\!-\!t^{2}\!K^{a_{2n-1}}\!K^{a_{2n}}\!\!+\!\frac{s^{2}}{
\ell _{\!e\!f\!\!f}^{2}}e^{a_{2n-1}}e^{a_{2n}}\!\!\Big), \\
=n \int_{0}^{1} dt \int_{0}^{t} ds\,\epsilon
_{\!A_{\!1}...A_{2n\!+\!1}}\theta
^{\!A_{\!1}\!A_{2}}e^{\!A_{3}}\!\Big(
\!{\mathcal{R}}^{\!A_{4}\!A_{5}}\!\!+\!t^{2}\theta
_{\,\,\,C}^{\!A_{4}}\theta ^{C\!A_{5}}\!\!+\!\!\frac{s^{2}}{\ell
_{\!e\!f\!\!f}^{2}}\!e^{\!A_{4}}e^{\!A_{5}}\!\!\Big)\times\cdots
\nonumber\\
\cdots\times\Big(\!
{\mathcal{R}}^{\!A_{2n}\!A_{2n+1}}\!\!+\!t^{2}\theta
_{\,\,\,\,\,\,F}^{A_{2n}}\theta
^{F\!A_{2n+1}}\!\!+\!\frac{s^{2}}{\ell
_{\!e\!f\!\!f}^{2}}e^{\!A_{2n}}e^{\!A_{2n\!+\!1}}\!\!\Big),
\end{eqnarray}
where the extrinsic curvature $K^{A}\!=\!K_{B}^{A}e^{B}$ satisfies
$K_{AB}\!=\!-h_{A}^{C}h_{B}^{D}n_{C;D}$, with $n^{A}$ the outward
unit normal vector at the boundary. The orthonormal frame takes the
block-diagonal form $e^{1}\!=\!Nd\rho $,
$e^{a}\!=\!e_{\,\,i}^{a}dx^{i}$, such that the only non-vanishing
components of $\theta^{AB}$ are $\theta
^{1a}\!=\!K^{a}\!=\!K_{j}^{i}e_{\,\,i}^{a}dx^{j}$, and the
submanifold Levi-Civita tensor is $\epsilon
_{a_{\!1}...a_{d}}\!=\!\epsilon _{1a_{\!1}...a_{d}}$.
${\mathcal{R}}^{AB}$ is the intrinsic curvature 2-form, that for a
radial foliation contains only components on the boundary
submanifold. Remarkably, the form of $B_{2n}$ is preserved
regardless the particular theory considered, only the corresponding
coupling constant changes accordingly, as shown below.
\subsection{Variational principle and boundary conditions}

An arbitrary variation of the action produces the equations of motion plus
contributions to the surface term that can be traced back to the bulk and
boundary terms in (\ref{lovelock})
\begin{eqnarray}
\delta I_{2n+1} =\int_{\!M}\!(E.O.M.)\!+\frac{1}{8\pi G_{\!D}}
\!\int_{\!\partial \!M}\sum_{p=1}^{n}\frac{p\alpha
_{p}}{(D\!-\!2p)!} \,\epsilon _{\!a_{\!1}\!...\!a_{2n}}\!\delta
\!K^{a_{1}}\hat{{\mathcal{R}}}
^{a_{2}a_{3}}...\hat{{\mathcal{R}}}^{a_{2p\!-\!2}a_{2p\!-\!1}}e^{a_{2p}}
\!...e^{a_{2n}}  \nonumber \\
 + 2nc_{2n}\!\int_{\!\partial
\!M}\!\int_{0}^{1}\!\!dt\,\epsilon _{a_{1}\!...\!a_{2n}}\!\delta
\!K^{a_{1}}e^{a_{2}}\Big(\!\hat{{\mathcal{R}}}^{a_{3}a_{4}}\!+\!\frac{t^{2}}{
\ell
_{\!e\!f\!\!f}^{2}}\!e^{a_{3}}e^{a_{4}}\!\!\Big)\!...\!\Big(\!\hat{
{\mathcal{R}}}^{a_{2n\!-\!1}a_{2n}}\!+\!\frac{t^{2}}{\ell
_{\!e\!f\!\!f}^{2}}
e^{\!a_{2n\!-\!1}}e^{a_{2\!n}}\!\!\Big)  \nonumber \\
 -2nc_{2\!n}\!\!\int_{\!\partial
\!M}\!\int_{0}^{1}\!\!\!dt\,t\,\epsilon
_{\!a_{\!1}\!...\!a_{\!2\!n}}\!(\!\delta
\!K^{\!a_{\!1}}e^{a_{2}}\!\!-\!\!K^{\!a_{\!1}}\delta
e^{a_{2}}\!)\Big(
\!\!{\mathcal{R}}^{\!a_{3}a_{4}}\!\!-\!t^{2}\!K^{\!a_{3}}\!K^{\!a_{4}}\!
\!+\!\frac{t^{2}}{\ell _{\!e\!f\!\!f}^{2}}e^{\!a_{3}}e^{\!a_{4}}\!\!
\Big)\!\times\cdots
\nonumber\\
\cdots\times\!\Big(\!\!{\mathcal{R}}^{\!a_{\!2n\!-\!1}a_{\!2n}}\!\!-\!t^{2}\!K^{\!a_{
\!2n\!-\!1}}\!K^{\!a_{\!2n}}\!\!+\!\!\frac{t^{2}}{\ell
_{\!e\!f\!\!f}^{2}} e^{a_{\!2n\!-\!1}}e^{a_{\!2n}}\!\!\Big)\!.
\label{var2n+1}
\end{eqnarray}

Here, we have extensively used the Gauss-Coddazzi relation for the boundary
components of the Riemann 2-form%
\begin{equation}
\hat{{\mathcal{R}}}^{ab}={\mathcal{R}}^{ab}\!-\!K^{a} K^{b},
\end{equation}
that in the standard tensorial notation reads
\begin{equation}
\hat{R}_{ij}^{kl}=R_{ij}^{kl}-K_{i}^{k}K_{j}^{l}+K_{j}^{k}K_{i}^{l}.
\end{equation}
A well-defined action principle for Lovelock-AdS gravity amounts to
the on-shell cancellation of the surface term in eq.(\ref{var2n+1})
by imposing suitable boundary conditions, that either are derived
from, or at least, are compatible with the asymptotic behavior of
the metric (\ref{radial})-(\ref{gFG}).

For the extrinsic curvature, the FG expansion produces
\begin{equation}
\!\!\!\!\!\!\!\!\!\!\!\!\!\!\!K_{j}^{i}=h^{ik}K_{kj}=\!\frac{1}{\ell
_{\!e\!f\!\!f}}\delta _{j}^{i}-\frac{\rho }{\ell _{\!e\!f\!\!f}}
(g_{(0)}^{-1}g_{(1)})_{j}^{i}-\frac{\rho ^{2}}{\ell _{\!e\!f\!\!f}}
(2g_{(0)}^{-1}g_{(2)}\!-
\!g_{(0)}^{-1}g_{(1)}g_{(0)}^{-1}g_{(1)})_{j}^{i}+...\,, \label{KFG}
\end{equation}
where the indices at the r.h.s. of the above equation are raised
with the conformal structure $g_{(0)}^{ij}$. Then, the extrinsic
curvature on the boundary is finite
\begin{equation}
K_{j}^{i}=\frac{1}{\ell _{\!e\!f\!\!f}}\delta _{j}^{i}.  \label{K=delta}
\end{equation}
In any gravity theory, $h_{ij}$ and $K_{ij}$ are independent
variables, because the extrinsic curvature defines the canonical
momentum $\pi ^{ij}$. The fact that the extrinsic curvature can be
written in terms of the coefficients $g_{(k)}$ in the expansion of
the metric does not mean that it is determined only by the metric
$g_{(0)}$. Indeed, as it is well-known, not even $h_{ij}$ is
completely determined by solving the second-order field equations
with only $g_{(0)}$ as the initial data (Fefferman-Graham ambiguity
for the coefficient $g_{(n)}$ with $n\!=\![D/2]$). Then, $K_{ij}$
remains as an independent variable even though the first terms in
the expansion (\ref{KFG}) are fixed by $g_{(0)}$.

For the variational problem in odd dimensions, we will consider that at the
boundary the variations obey
\begin{equation}
\delta K_{j}^{i}=0,  \label{deltaK=0}
\end{equation}
that is a regular boundary condition compatible with fixing the
conformal metric $g_{(0)}$ on $\partial M$ \cite{miol,OleaKounter}.
Therefore, this boundary condition does not spoil the AdS/CFT
interpretation of the conformal structure $g_{(0)}$ as a given data
for the holographic reconstruction of the spacetime in the gravity
side, and whose dual CFT on the boundary does not have gravitational
degrees of freedom.

The last line in (\ref{var2n+1}) is identically canceled by the
conditions (\ref{K=delta}), (\ref{deltaK=0}). The asymptotic
behavior (\ref{AAdSR}) for the curvature determines the coupling
constant $c_{2n}$
\begin{equation}
c_{2n}=\frac{1}{16\pi
nG_{\!D}}\Big[\!\int_{0}^{1}\!\!\!dt\,(t^{2}\!-\!1)^{n-\!1}\!
\Big]^{\!-1}\,\sum_{p=1}^{n}\frac{(-1)^{p}p\,\alpha
_{p}}{(D\!-\!2p)!}\ell _{\!e\!f\!\!f}^{2(\!n-p)},  \label{c2n}
\end{equation}
in order to cancel the rest of the surface term (\ref{var2n+1}).

In the standard Dirichlet regularization for AdS gravity, fixing the
conformal structure $g_{(0)ij}$ in the boundary metric (\ref{hFG}),
(\ref{gFG}) will require the addition of counterterms to cancel the
divergence at the boundary $\rho =0$ \cite{Papadimitriou-Skenderis}.
In our case, we select the boundary conditions (\ref {AAdSR}),
(\ref{K=delta}) and (\ref{deltaK=0}), which are regular on the
asymptotic region, such that the regularization process is encoded
in the boundary terms already present and there is no need of
further addition of counterterms.

\subsection{Conserved quantities and vacuum energy.}

The Noether theorem applied to Lovelock-AdS gravity states that
there is a set of conserved charges $Q(\xi)$ associated to
asymptotic Killing vectors $ \xi $, that are defined as
$(D\!-\!2)-$forms, and therefore, are integrated on the boundary of
a spatial section at constant time. More precisely, we take a
timelike ADM foliation for the line element at the boundary
\begin{equation}
h_{ij}\,dx^{i}dx^{j}\!=\!-\tilde{N}^{2}(t)\,dt^{2}\!+\!\sigma
_{\underline{m}\,\underline{n}}( d\varphi
^{\underline{m}}+\!\tilde{N}^{\underline{m}}\,dt) ( d\varphi
^{\underline{n}}+\!\tilde{N}^{\underline{n}}\,dt) \,,
\end{equation}
with the coordinates $x^{i}=\left( t,\varphi
^{\underline{m}}\right) $, that is defined by the unit normal
vector $u_{i}=(-\tilde{N},\vec{0})$. The charges are then given as
the integration on the boundary $\Sigma$ of a spatial section,
parameterized by $\varphi ^{\underline{m}}$
\begin{equation}
Q(\xi )\!=\!\int_{\!\Sigma }\!d^{D-\!2}\varphi \,\sqrt{\sigma
}\,u_{j} Q_{i}^{j}\,\xi ^{i}\,.  \label{Q}
\end{equation}
In the above formula, $\sigma $ denotes the determinant of the
metric $ \sigma _{\underline{m}\,\underline{n}}$, related to $h$ by
$\sqrt{\!-h}\!=\!\tilde{N} \sqrt{\sigma }$, and $\xi ^{i}$ is an
asymptotic Killing vector. In odd dimensions, the expression for the
integrand appears naturally split in two pieces
\begin{equation}
Q_{i}^{j}=q_{i}^{j}+q_{(0)i}^{j}\,, \label{Qintegrand}
\end{equation}
with
\begin{eqnarray}
&&\!\!\!\! q_{i}^{j} =\frac{1}{2^{n-2}}\,\delta _{\lbrack
i_{1}i_{2}...i_{2n}]}^{[jj_{2}...j_{2n}]}\,K_{i}^{i_{1}}\delta
_{j_{2}}^{i_{2}}\left[\!\frac{1}{16\pi
G_{\!D}}\sum_{p=1}^{n}\frac{p\alpha _{p}
}{(D\!-\!2p)!}\hat{R}_{j_{3}j_{4}}^{i_{3}i_{4}}...\hat{R}
_{j_{2p-1}j_{2p}}^{i_{2p-1}i_{2p}}\delta _{\lbrack
j_{2p+1}j_{2p+2}]}^{[i_{2p+1}i_{2p+2}]}...\delta _{\lbrack
j_{2n-1}j_{2n}]}^{[i_{2n-1}i_{2n}]}\right.  \nonumber \\
&&\,\,\,\,\,\,\,\,\,\,\,\,\,\,\,\,\,\,\,\,\,\,\,\,\,\,\,\,\,\,\,\,\,\,\,\,\,\,\,\,\,\,\,\,
\,\,\,\,\,\,\,\,\,\,\,\,\,\,\,\,\,\,\,\,\,\,\,\,\,\,\left.
+nc_{2n}\!\int_{0}^{1}\!\!\!dt\,\Big(\!\hat{R}_{j_{3}j_{4}}^{i_{3}i_{4}}\!+\!
\frac{t^{2}}{\ell _{\!e\!f\!\!f}^{2}}\delta _{\lbrack
j_{3}j_{4}]}^{[i_{3}i_{4}]}
\Big)...\Big(\!\hat{R}_{j_{2n-1}j_{2n}}^{i_{2n-1}i_{2n}}\!+\!\frac{t^{2}}{\ell
_{\!e\!f\!\!f}^{2}}\delta _{\lbrack
j_{2n-1}j_{2n}]}^{[i_{2n-1}i_{2n}]}\Big)\!\right] \label{qij},
\\
&&\!\!\!\!\!\!\!q_{(0)i}^{j}\!=\!-\frac{nc_{2n}}{2^{n-2}}\!\int_{0}^{1}\!\!dt\,t\,\delta
_{\lbrack i_{1}i_{2}...i_{2n}]}^{[jj_{2}...j_{2n}]}(\delta
_{j_{2}}^{i_{2}}\!K_{i}^{i_{1}}\!+\!\delta
_{i}^{i_{2}}\!K_{j_{2}}^{i_{1}}\!)
\Big(\!R_{j_{3}j_{4}}^{i_{3}i_{4}}\!-\!t^{2}K_{[j_{3}j_{4}]}^{[i_{3}i_{4}]}\!+\!\frac{
t^{2}}{\ell _{\!e\!f\!\!f}^{2}}\delta _{\lbrack
j_{3}j_{4}]}^{[i_{3}i_{4}]}\Big)\times \cdots \nonumber
\\
&& \hspace{50mm}\cdots\times\Big(\!
R_{j_{2n\!-\!1}j_{2n}}^{i_{2n\!-\!1}i_{2n}}\!-\!t^{2}K_{[j_{2n\!-\!1}j_{2n}]}^{[i_{2n\!-\!1}i_{2n}]}\!+\!
\frac{t^{2}}{\ell _{\!e\!f\!\!f}^{2}}\delta _{\lbrack
j_{2n\!-\!1}j_{2n}]}^{[i_{2n\!-\!1}i_{2n}]}\Big), \label{qzeroij}
\end{eqnarray}
where we have used the shorthand
$K_{[jl]}^{[ik]}=K_{j}^{i}K_{l}^{k}-K_{l}^{i}K_{j}^{k}$.

Equation (\ref{Qintegrand}) defines a splitting of the charges
\begin{equation}
Q(\xi)=q(\xi )+q_{0}(\xi )\,, \label{splitting}\end{equation} where
\begin{equation}
q(\xi)\!=\!\int_{\!\Sigma }\!\!d^{D-2}\varphi \sqrt{\sigma }
\,u_{j}q_{i}^{j}\,\xi ^{i}  \label{qxi}
\end{equation}
will provide the mass and angular momentum for AAdS black hole
solutions in Lovelock gravity. It can be shown that eq.(\ref{qij})
can be factorized in any odd dimension as
\begin{equation}
q_{i}^{j}=\frac{1}{2^{n-2}}\delta _{\lbrack
i_{1}i_{2}...i_{2n}]}^{[jj_{2}...j_{2n}]}K_{i}^{i_{1}}\delta
_{j_{2}}^{i_{2}}
\Big(\!\hat{R}_{j_{3}j_{4}}^{i_{3}i_{4}}\!+\!\frac{1}{\ell_{\!e\!f\!\!f}
^{2}}\delta _{\lbrack j_{3}j_{4}]}^{[i_{3}i_{4}]}\Big){\mathcal{P}}
_{j_{5}...j_{2n}}^{i_{5}...i_{2n}},  \label{qijfact}
\end{equation}
where ${\mathcal{P}}$ is a Lovelock-type polynomial of
$(n-2)-$degree in the Riemann tensor $\hat{R}_{kl}^{ij}$ and the
antisymmetrized Kronecker delta $ \delta _{\lbrack kl]}^{[ij]}$
\begin{equation}
{\mathcal{P}}_{j_{5}...j_{2n}}^{i_{5}...i_{2n}}=\sum_{p=0}^{n-2}\Big(\!nc_{2n}
\frac{D_{p}}{\ell _{\!e\!f\!\!f}^{2p}}\!+\!\frac{F_{p}}{16\pi
G_{\!D}}\!\Big)\hat{R} _{j_{5}j_{6}}^{i_{5}i_{6}}...\hat{R}
_{j_{2(n-p)-1}j_{2(n-p)}}^{i_{2(n-p)-1}i_{2(n-p)}}\delta _{\lbrack
j_{2(n-p)+1}j_{2(n-p+1)}]}^{[i_{2(n-p)+1}i_{2(n-p+1)}]}...\delta
_{\lbrack j_{2n-1}j_{2n}]}^{[i_{2n-1}i_{2n}]},
\end{equation}
with the coefficients of the expansion given by
\begin{equation}
D_{p}=\sum_{q=0}^{p}\frac{(\!-1\!)^{p-q}}{2q\!+\!1}\Big(\!\!
\begin{array}{c}
n\!-\!1 \\
q
\end{array}
\!\!\Big)\,\,\,\,\,\,\,\,,\!\!\!\!\!\!\qquad
F_{p}=\sum_{q=0}^{p}\frac{(\!-1\!)^{p-q}\,(n\!-\!q)\,\alpha _{n-q}}{
(2q\!+\!1)!\,\,\ell _{\!e\!f\!\!f}^{2(p-q)}}.
\end{equation}

As the tensorial combination $\hat{R}_{ij}^{kl}\!+\!\frac{1}{\ell
_{\!e\!f\!\!f}^{2}} \delta _{\lbrack ij]}^{[kl]}$ is a part of the
curvature of the AdS group with an effective radius $\ell
_{\!e\!f\!\!f}$, the factorization (\ref{qijfact}) implies that the
charge $q(\xi )$ vanishes identically for global AdS spacetime. Note
that for Einstein-Hilbert-AdS gravity, $F_{p}=0$ and the above
expression for $q(\xi )$ recovers the corresponding charge in \cite
{OleaKounter}. As a consequence, the quantity $q_{0}(\xi )$
\begin{equation}
q_{0}(\xi )\!=\!\int_{\!\Sigma }\!\!d^{D-2}\varphi \,\sqrt{\sigma }
\,u_{j}q_{(0)i}^{j}\,\xi ^{i}\,,  \label{q0xi}
\end{equation}
is truly a tensorial formula for the vacuum energy for AAdS spacetimes in
Lovelock gravity, inexistent in previous literature.

A static black hole solution for Lovelock gravity (\ref{lovelock})
for both odd and even dimensions $D$ is given by the metric
\begin{equation}
ds^{2}=-\Delta ^{2}(r)dt^{2}+\frac{dr^{2}}{\Delta
^{2}(r)}+r^{2}\gamma_{ \underline{m}\,\underline{n}}d\varphi
^{\underline{m}}d\varphi ^{\underline{n} },  \label{ds2BH}
\end{equation}
with $\Delta (r)$ given by
\begin{equation}
\sum_{p=1}^{[(\!D-\!1\!)\!/2]}\!\frac{\alpha
_{p}}{(D\!-\!2p\!-\!1)!}\Big(\!\frac{k\!-\!\Delta
^{2}}{r^{2}}\!\Big)^{p}=\frac{2\Lambda}{(D\!-\!1)!}+\frac{\mu
}{(D\!-\!3)!\,r^{D-\!1}}, \label{ko}
\end{equation}
where $\mu $ appears as an integration constant in the first integral of the
$rr$ component of the Lovelock equations of motion (\ref{EOMlovelock})
\begin{equation}
\frac{(\Delta ^{2})^{\prime
}}{r}\sum_{p=1}^{[(\!D-\!1\!)\!/2]}\frac{p\alpha _{p}}{
(D\!-\!2p\!-\!1)!}\Big(\!\frac{k\!-\!\Delta
^{2}}{r^{2}}\!\Big)^{p-1}-\sum_{p=1}^{[(\!D-\!2)
\!/2]}\!\frac{\alpha
_{p}}{(D\!-\!2p\!-\!2)!}\Big(\!\frac{k\!-\!\Delta
^{2}}{r^{2}}\!\Big) ^{p}=-\frac{2\Lambda}{(D\!-\!2)!}\,, \label{rr}
\end{equation}
and the prime stands for the derivative with respect to $r$
coordinate. The metric $
\gamma_{\underline{m}\,\underline{n}}$ ($\underline{m},\underline{n}%
=1,...,D-2$) defines the line element of the transversal section $\Sigma
_{D-2}^{k}$ whose curvature is a constant $k=\pm 1,0$. Black hole solution (%
\ref{ds2BH}) possesses an event horizon $r_{+}$, which is the largest root
of the equation $\Delta ^{2}(r_{+})=0$. For this configuration, the only
nonvanishing components of the extrinsic curvature are
\begin{equation}
K_{t}^{t}=-\Delta ^{\prime }\,\,\,\,\,\,\,\,\,\,,\!\!\!\!\!\qquad
K_{\underline{m}}^{\underline{n}}=-\frac{ \Delta }{r}\delta
_{\underline{m}}^{\underline{n}}\,,  \label{KSch}
\end{equation}
whereas the intrinsic curvature is
\begin{equation}
R_{\underline{m}_{2}\underline{n}_{2}}^{\underline{m}_{1}\underline{n}_{1}}=
\frac{k}{r^{2}}\delta _{[
\underline{m}_{2}\underline{n}_{2}]}^{[\underline{
m}_{1}\underline{n}_{1}]}\,,  \label{Rsphere}
\end{equation}
which, in turn, produces the boundary components of the Riemann
tensor to be
\begin{equation}
\hat{R}^{t\underline{n}}_{t\underline{m}}\!=\!-\frac{(\Delta
^{2})^{\prime}}{2r}\delta
^{\underline{n}}_{\underline{m}}\qquad,\,\,\,\,\,\,\,\,
\hat{R}_{\underline{m}_{2}\underline{n}_{2}}^{\underline{m}_{1}\underline{n}
_{1}}\!=\!\frac{k\!-\!\Delta ^{2}}{r^{2}}\delta _{\lbrack
\underline{m}_{2}
\underline{n}_{2}]}^{[\underline{m}_{1}\underline{n}_{1}]}\,.
\label{hatRBH}
\end{equation}

Differentiating eq.(\ref{ko}) with respect to the horizon radius
$r_{+}$ and combining eqs.(\ref{ko}), (\ref{rr}), we obtain the
relation
\begin{equation}
\frac{\partial {\mu }}{\partial {r_{+}}}=(D\!-\!3)!\,(\Delta
^{2})^{\prime
}|_{r_{\!+}}\,\sum_{p=1}^{[(\!D-\!1\!)\!/2]}\!\frac{p\alpha
_{p}}{(D\!-\!2p\!-\!1)!} \,r_{+}^{D-2p-1}k^{p-1}.  \label{dmudrplus}
\end{equation}
From the equation (\ref{ko}) that dictates the form of the function
$\Delta ^{2}(r)$ in the metric, we can obtain the asymptotic
behavior ($r\rightarrow \infty $)
\begin{equation}
\Delta ^{2}(r)\approx k+\frac{r^{2}}{\ell
_{\!e\!f\!\!f}^{2}}-\frac{\mu }{(D\!-\!3)!} \left[
\sum_{p=1}^{[(\!D-\!1\!)\!/2]}\!\frac{p\alpha _{p}(-\ell
_{\!e\!f\!\!f}^{-2})^{p-1}}{(D\!-\!2p\!-\!1)!}\right]
^{-1}\!\frac{1}{r^{D-3}}+... \label{Delta2asympt}
\end{equation}

The corresponding {\textit{mass}} is given by the evaluation of
eq.(\ref{qxi}) for the timelike Killing vector $\xi =\partial _{t}$
\begin{equation}
q(\partial _{t})\!=\!M\!=\!(D\!-\!2)!\,vol(\Sigma
_{D-\!2}^{k})(\Delta ^{2})^{\prime }\Big[ \frac{1}{16\pi
G_{\!D}}\sum_{p=1}^{n}\frac{p\alpha _{p}}{(D\!-\!2p)!}
r^{D-2p}(k-\Delta
^{2})^{p-1}+nc_{2n}r\int_{0}^{1}\!\!dt\Big(\!k-\Delta ^{2}+\frac{
t^{2}r^{2}}{\ell
_{\!e\!f\!\!f}^{2}}\!\Big)^{\!n-\!1}\Big]\!\Big|^{\!\infty }.
\label{MDeltaodd}
\end{equation}
Using the asymptotic form of $\Delta ^{2}(r)$ from
(\ref{Delta2asympt}), we see that the divergent terms $O(r^{D-1})$
in the evaluation of the above formula exactly cancel out and one
gets the finite result
\begin{equation}
M\!=\!\frac{(D\!-\!2)\,vol(\Sigma _{D-\!2}^{k})}{16\pi G_{\!D}}\mu .
\label{massodd}
\end{equation}
The zero-point ({\textit{vacuum)}} energy is then given by
(\ref{q0xi}) as
\begin{equation}
q_{0}(\partial _{t})=E_{0}=2nc_{2n}(D\!-\!2)!\,vol(\Sigma
_{D-\!2}^{k})\Big(\!\Delta ^{2}\!-\!\frac{r(\Delta ^{2})^{\prime
}}{2}\!\Big)\int_{0}^{1}\!\!\!dt\,t\,\Big(\!k\!-\!t^{2}\Delta
^{2}\!+\!\frac{t^{2}r^{2}}{\ell
_{\!e\!f\!\!f}^{2}}\!\Big)^{\!n-1}\Big| ^{\infty },  \label{vac}
\end{equation}
that can be worked out using the asymptotic form
(\ref{Delta2asympt}), giving the finite result
\begin{equation}
E_{0}=(-k)^{n}\frac{vol(\Sigma _{D-\!2}^{k})}{16\pi
nG_{\!D}}(2n\!-\!1)!!^{2} \sum_{p=1}^{n}\frac{(-1)^{p-1}p\alpha
_{p}}{(D\!-\!2p)!}\ell _{\!e\!f\!\!f}^{2n-2p}. \label{Ezero}
\end{equation}

\subsection{Black Hole Entropy.}

The Euclidean period $\beta $ is defined as the inverse of black hole
temperature $T$ such that in the Euclidean sector the solution (\ref{ds2BH})
does not have a conical singularity at the horizon. In doing so, one obtains
$\beta =4\pi /(\Delta ^{2})^{\prime }|_{r_{+}}$. The black hole entropy $S$
is defined in the canonical ensemble (the surface gravity is kept fixed at
the horizon) as
\begin{equation}
S=I^{E}+\beta \mathcal{E},  \label{Sdef}
\end{equation}
in terms of the total Euclidean action $I^{E}$ and the thermodynamical
energy
\begin{equation}
\mathcal{E}\!=\!-\frac{\partial I^{E}}{\partial \beta }  \label{TEnergydef}
\end{equation}
of the black hole. The Euclidean bulk action is evaluated for a
static black hole of the form (\ref{ds2BH}) as
\begin{equation}
I_{bulk}^{E}=-\frac{(D\!-\!2)!}{16\pi G_{\!D}}vol(\Sigma
_{D-\!2}^{k})\,\beta \sum_{p=1}^{n}\frac{p\alpha
_{p}}{(D\!-\!2p)!}[r^{D-2p}(\Delta ^{2})^{\prime }\,(k\!-\!\Delta
^{2})^{p-1}]|_{r_{+}}^{\infty },  \label{IEbulkBH}
\end{equation}
and it is rendered finite by the addition of the suitable boundary
term (\ref {B2ntensor}), whose evaluation in the Euclidean solution
is
\begin{equation}
\int_{\!\partial {\!M}}\!\!B_{2n}^{E}\!=\!-n(D-2)!\,vol(\Sigma
_{D-2}^{k})\beta \Big[r(\Delta ^{2})^{\prime
}\!\int_{0}^{1}\!\!\!dt\Big(\!k-\Delta ^{2}+\frac{ t^{2}r^{2}}{\ell
_{\!e\!f\!\!f}^{2}}\!\Big)^{\!n-1}\!+2\Big(\!\Delta
^{2}\!-\frac{r(\Delta ^{2})^{\prime
}}{2}\!\Big)\int_{0}^{1}\!\!\!dt\,t\,\Big(\!k-t^{2}\Delta
^{2}+\frac{ t^{2}r^{2}}{\ell
_{\!e\!f\!\!f}^{2}}\!\Big)^{\!n-1}\Big]\Big|^{\infty }.
\end{equation}
Therefore, the total action contains two pieces. At $r=\infty $, the
contribution from the bulk action $I_{bulk}^{E}$ combines with the
boundary term $c_{2n}\int_{\!\partial {\!M}}\!B_{2n}^{E}$ to produce
$-\beta $ times the Noether charge $Q(\partial _{t})=M+E_{0}$
\begin{equation}
I_{2n+1}^{E}=\frac{(D\!-\!2)!}{16\pi G_{\!D}}vol(\Sigma
_{D-\!2}^{k})\Big[4\pi \sum_{p=1}^{n}\frac{p\alpha
_{p}}{(D\!-\!2p)!}r_{+}^{D-2p}k^{p-1}-\frac{\beta \mu
}{(D\!-\!3)!}-\beta
(\!-k\!)^{\!n}\frac{(2n\!-\!1\!)!!^{2}}{n(D\!-\!2)!}
\sum_{p=1}^{n}\frac{(-1)^{p-1}p\alpha _{p}}{(D\!-\!2p)!}\ell
_{\!e\!f\!\!f}^{2n-2p} \Big].
\end{equation}
This identification guarantees that all the divergencies at radial
infinity are exactly canceled.

The definition of thermodynamic energy, using equation (\ref{dmudrplus}),
gives
\begin{equation}
{\mathcal{E}}=-\frac{\partial I_{\!2n+1}^{E}/\partial
r_{+}}{\partial \beta /\partial r_{+}}=M+E_{0},
\end{equation}
which recovers the same total energy as from the Noether charge
$Q(\partial _{t})$ of (\ref{splitting}). As a consequence, the
entropy (\ref{Sdef}) is simply given by the Noether charge evaluated
at the horizon
\begin{equation}
S=\frac{(D\!-\!2)!}{4G_{\!D}}vol(\Sigma
_{D-\!2}^{k})\sum_{p=1}^{n}\frac{p\alpha _{p}
}{(D\!-\!2p)!}r_{+}^{D-2p}k^{p-1}. \label{entroodd}
\end{equation}

\section{$D=2n$ dimensions}

For even dimensions, an alternative regularization procedure was
developed originally for Einstein-Hilbert-AdS action in
\cite{OleaJHEP} and applied to the same problem in AAdS gravity in
Einstein-Gauss-Bonnet theory \cite{Kofinas-Olea}. As we shall
explicitly show below, the universal form of the boundary term
that renders the conserved charges and Euclidean action finite in
Lovelock-AdS gravity in $D=2n$ corresponds to the (maximal) $n-$th
Chern form \cite{chern,eguchi,nakahara}
\begin{equation}
B_{2n-1}=n\!\int_{0}^{1}\!\!dt\,\epsilon _{A_{1}...A_{2n}}\theta
^{A_{1}A_{2}}( {\mathcal{R}}^{A_{3}A_{4}}\!+\!t^{2}\theta
_{\,\,\,\,C}^{A_{3}}\theta
^{CA_{4}})...({\mathcal{R}}^{\!A_{2n\!-\!1}\!A_{2n}}\!+\!t^{2}\theta
_{\,\,\,\,\,\,\,\,\,\,\,F}^{\!A_{2n\!-\!1}}\theta ^{F\!A_{2n}}).
\label{b1}
\end{equation}
Eq.(\ref{b1}) can be projected to the boundary indices to work out
its equivalence in tensorial notation
\begin{eqnarray}
B_{2n-1} \!\!\!\!\!&=&\!\!\!\!\!2n\!\int_{0}^{1}\!\!dt\,\epsilon
_{a_{1}\!...\!a_{2n\!-\!1}}K^{a_{1}}({\mathcal{R}}^{a_{2}a_{3}}\!-
\!t^{2}K^{a_{2}}\!K^{a_{3}})...({\mathcal{R}}^{a_{2n\!-\!2}a_{2n\!-\!1}}\!-
\!t^{2}\!K^{a_{2n\!-\!2}}\!K^{a_{2n\!-\!1}})  \label{b2} \\
&\!\!\!\!\!\!=\!\!\!\!\!\!&2n\sqrt{\!-h}\!\int_{0}^{1}\!\!\!dt\delta
_{\lbrack
i_{1}...i_{2n\!-\!1}]}^{[j_{1}...j_{2n\!-\!1}]}K_{j_{1}}^{i_{1}}\Big(\frac{1
}{2}R_{j_{2}j_{3}}^{i_{2}i_{3}}\!-\!t^{2}K_{j_{2}}^{i_{2}}K_{j_{3}}^{i_{3}}\!
\Big)...\Big(\frac{1}{2}R_{j_{2n\!-\!2}j_{2n\!-\!1}}^{i_{2n\!-\!2}i_{2n\!-
\!1}}\!-\!t^{2}K_{j_{2n-2}}^{i_{2n-2}}K_{j_{2n-1}}^{i_{2n-1}}\!\Big)
d^{2n\!-\!1}\!x.  \label{b3}
\end{eqnarray}
In the last formula, the parametric integration reflects the action
of the Cartan homotopy operator, used to obtain the correction to
the Euler characteristic due to the introduction of a boundary in
the Euler theorem. The integral in $t$ is a convenient shorthand,
but it also generates the suitable coefficients in the binomial
expansion
\begin{equation}
B_{2n-1}=2n\sqrt{\!-h}\sum_{p=0}^{n-1}\frac{(-1)^{n-p-1}}{2^{p}\,(2n\!-\!2p\!-\!1)}
b_{2n-1}^{(p)},
\end{equation}
where
\begin{equation}
b_{2n-1}^{(p)}=\delta _{\lbrack j_{1}\cdots j_{2p}\cdots
j_{2n-1}]}^{[i_{1}\cdots i_{2p}\cdots
i_{2n-1}]}R_{i_{1}i_{2}}^{j_{1}j_{2}}\cdots
R_{i_{2p-1}i_{2p}}^{j_{2p-1}j_{2p}}\,K_{i_{2p+1}}^{j_{2p+1}}\cdots
K_{i_{2n-1}}^{j_{2n-1}}\,.
\end{equation}
The surface term coming from an arbitrary on-shell variation of the
action (\ref{lovelock}) adopts a slightly simpler form than in the
odd-dimensional case
\begin{eqnarray}
\delta I_{2n}=\int_{\!\partial\!M}\frac{1}{8\pi
G_{\!D}}\sum_{p=1}^{n-1}\!\frac{ p\alpha _{p}}{(D\!-\!2p)!}\epsilon
_{a_{1}...a_{2n\!-\!1}}\delta \!K^{a_{\!1}}\hat{
{\mathcal{R}}}^{a_{2}a_{3}}...\hat{{\mathcal{R}}}
^{a_{2p-2}a_{2p-\!1}}e^{a_{2p}}...e^{a_{2n-\!1}}\!+ \nonumber
\\
+2nc_{2n\!-\!1}\epsilon _{a_{\!1}...a_{2n\!-\!1}}\delta\!
K^{a_{\!1}}\hat{{\mathcal{R}}}^{a_{2}a_{3}}...\hat{
{\mathcal{R}}}^{a_{2n\!-\!2}a_{2n\!-\!1}}.  \label{deltaI2n}
\end{eqnarray}
An appropriate choice of the coupling constant $c_{2n-1}$ as
\begin{equation}
c_{2n-1}=-\frac{1}{16\pi nG_{\!D}}\sum_{p=1}^{n-1}\frac{p\alpha
_{p}}{(D\!-\!2p)!} (-\ell _{\!e\!f\!\!f}^{2})^{n-p}.
\end{equation}
makes the above expression vanish identically for AAdS spacetimes (\ref%
{AAdSR}). The regularity of the asymptotic condition (\ref{AAdSR}) implies
that the well-defined action principle achieved in this way is also a finite
one, because no additional divergences are induced by the addition of the
Kounterterms (\ref{b3}).

\subsection{Conserved Charges}

In Einstein-Hilbert and Einstein-Gauss-Bonnet with negative cosmological
constant, we have seen that the addition of boundary terms with explicit
dependence on the extrinsic curvature $K_{ij}$ solve at once two problems
that in general are not necessarily related: the variational principle and
the finiteness of the Noether charges and Euclidean action. Whenever the
action is stationary for boundary conditions compatible with the asymptotic
structure of AAdS spacetimes, the theory does not require a further
regularization on top of the addition of $B_{d}$ in eq.(\ref{lovelock}).

The conserved charges constructed using the Noether theorem have the
form
\begin{equation}
Q(\xi )=\int_{\!\Sigma }d^{D-2}\varphi \,\sqrt{\sigma }\,u_{j}
Q_{i}^{j}\,\xi ^{i}\,,  \label{Q2n}
\end{equation}
with the integrand given by
\begin{equation}
Q_{i}^{j}\!=\!\frac{1}{2^{n-2}}\delta _{\lbrack
i_{1}i_{2}...i_{2n\!-\!1}]}^{[jj_{2}...j_{2n\!-\!1}]}K_{i}^{i_{1}}\!\!\left[\!
\frac{1}{ 16\pi G_{\!D}}\sum_{p=1}^{n-1}\!\frac{p\alpha
_{p}}{(D\!-\!2p)!}\hat{R}
_{j_{2}j_{3}}^{i_{2}i_{3}}...\hat{R}_{j_{2p\!-\!2}j_{2p\!-\!1}}^{i_{2p\!-\!2}i_{2p\!-\!1}}
\delta _{\lbrack j_{2p}j_{2p+1}]}^{[i_{2p}i_{2p+1}]}...\delta
_{\lbrack
j_{2n\!-\!2}j_{2n\!-\!1}]}^{[i_{2n\!-\!2}i_{2n\!-\!1}]}\!+\!nc_{2n\!-\!1}\hat{R}
_{j_{2}j_{3}}^{i_{2}i_{3}}\!...\hat{R}_{j_{2n\!-\!2}j_{2n\!-\!1}}^{i_{2n\!-
\!2}i_{2n\!-\!1}}\!\right] \!\!.  \!\!\!\!\!\!\label{Qij2n}
\end{equation}
The {\textit{mass}} for Lovelock-AdS black holes (\ref{ds2BH}),
(\ref{ko}) comes from the above formula for the Killing vector $\xi
=\partial _{t}$, that is
\begin{equation}
Q(\partial _{t})=M=(D\!-\!2)!\,vol(\Sigma _{D-\!2}^{k})(\Delta
^{2})^{\prime }\Big[ \frac{1}{16\pi
G_{\!D}}\sum_{p=1}^{n-1}\frac{p\alpha _{p}}{(D\!-\!2p)!}
r^{D-2p}(k\!-\!\Delta ^{2})^{p-1}+nc_{2n-1}(k\!-\!\Delta
^{2})^{n-1}\Big]\Big| ^{\infty }. \label{mass2n}
\end{equation}
As in the odd-dimensional case, taking the asymptotic expansion of the
functions involved shows that the divergences at order $r^{D-1}$ coming both
from the bulk and boundary parts of the action are exactly canceled. Thus,
we obtain
\begin{equation}
M=\frac{(D\!-\!2)vol(\Sigma _{D-\!2}^{k})}{16\pi G_{\!D}}\mu .
\label{Mass2n}
\end{equation}

\subsection{Black Hole Entropy}

The Euclidean bulk action $I_{bulk}^{E}$ is still given by the
even-dimensional equivalence of equation (\ref{IEbulkBH})
\[
I_{bulk}^{E}=-\frac{(D\!-\!2)!}{16\pi G_{\!D}}vol(\Sigma
_{D-\!2}^{k})\,\beta \,\sum_{p=1}^{n-1}\frac{p\alpha
_{p}}{(D\!-\!2p)!}[r^{D-2p}\,(\Delta ^{2})^{\prime }\,(k\!-\!\Delta
^{2})^{p-1}]|_{r_{+}}^{\infty },
\]
while the Euclidean continuation of the boundary term takes the form
\begin{equation}
\int_{\!\partial \!M}\!\!B_{2n-1}^{E}=-n(D\!-\!2)!\,vol(\Sigma
_{D-\!2}^{k})\,\beta \,(\Delta ^{2})^{\prime }\,(k\!-\!\Delta
^{2})^{n-1}\Big|^{\infty }\!. \label{B2n-1BH}
\end{equation}
In the total Euclidean action in even dimensions evaluated for a
black hole (\ref{ds2BH}), (\ref{ko})
\begin{equation}
I_{2n}^{E}=I_{bulk}^{E}+c_{2n-1}\!\int_{\partial
\!M}\!\!B_{2n-1}^{E}, \label{IE2n}
\end{equation}
the term at infinity corresponds to $-\beta M$, where $M$ is the
Noether mass in eqs. (\ref{mass2n}), (\ref{Mass2n}), that is
\begin{equation}
I_{2n}^{E}=\frac{(D\!-\!2)!}{16\pi G_{\!D}}vol(\Sigma
_{D-\!2}^{k})\Big[4\pi \sum_{p=1}^{n-1}\frac{p\alpha
_{p}}{(D\!-\!2p)!}r_{+}^{D-2p}k^{p-1}-\frac{\beta \mu
}{(D\!-\!3)!}\Big].
\end{equation}
The consistency between the regularization procedure and the
thermodynamic ensemble is corroborated by the fact that the
thermodynamic energy
\begin{equation}
{\mathcal{E}}=-\frac{\partial I_{2n}^{E}}{\partial \beta }=M,
\label{ftou}
\end{equation}
reobtains the corresponding Noether charge. Finally, the black hole
entropy is expressed in terms of the horizon $r_{+}$ in a similar
form as eq.(\ref {entroodd}) for the odd-dimensional case
\begin{equation}
S=\frac{(D\!-\!2)!}{4G_{\!D\!}}\,vol(\Sigma
_{D-\!2}^{k})\sum_{p=1}^{n-1}\frac{p\alpha
_{p}}{(D\!-\!2p)!}r_{+}^{D-2p}k^{p-1}.  \label{S_even}
\end{equation}

\section{Particular Cases}

\textbf{Einstein-Gauss-Bonnet-AdS Gravity.} In this case, all the
coefficients in the Lovelock series are vanishing but $\alpha _{0}=-2\Lambda
$, $\alpha _{1}=1$ and $\alpha _{2}=\alpha $, where $\alpha $ is an
arbitrary positive coupling constant. The effective AdS radius is modified
by the Gauss-Bonnet coupling as
\begin{equation}
\frac{1}{\ell _{\!e\!f\!\!f}^{2}}=\frac{1\pm
\sqrt{1-4(D\!-\!3)(D\!-\!4)\alpha /\ell ^{2}}}{
2(D\!-\!3)(D\!-\!4)\alpha }\,, \label{elleffGB}
\end{equation}
such that the solutions tend asymptotically to a constant curvature
spacetime with that radius. The Noether charge in the corresponding
odd and even dimensions, evaluated for a timelike Killing vector
$\xi =\partial _{t}$ for Boulware-Deser black holes
\begin{equation}
\Delta ^{2}(r)=k+\frac{r^{2}}{2(D\!-\!3)(D\!-4)\alpha }\Big[1\pm
\sqrt{1-\frac{ 4(D\!-\!3)(D\!-\!4)\alpha }{\ell
^{2}}+\frac{4(D\!-\!3)(D\!-\!4)\alpha \,\mu }{r^{D-1}}}\Big] ,
\label{BDBH}
\end{equation}
recovers the mass obtained by background-dependent methods
\cite{des-tek, padilla, der-kat-ogu, petrov} $\!$
\begin{equation}
M=\frac{(D\!-\!2)\,vol(\Sigma _{D-\!2}^{k})}{16\pi G_{\!D}}\mu .
\label{massBDBH}
\end{equation}
However, background-independent methods are the only ones that can
detect the presence of a vacuum energy for Einstein-Gauss-Bonnet
theory. The Dirichlet regularization for arbitrary couplings of
quadratic curvature terms, and therefore, useful to treat the EGB
action, is only known in five dimensions \cite{cvetic} and, for the
Gauss-Bonnet case, it has shown to be ambiguous \cite{ambig}. The
procedure carried out here reproduces, by direct replacement of the
corresponding Lovelock coefficients $\{\alpha _{0},\alpha
_{1},\alpha _{2}\}$ in eq.(\ref{Ezero}), the general formula for the
vacuum energy for EGB-AdS theory
\begin{equation}
E_{0}=(-k)^{n}\frac{vol(\Sigma _{D-\!2}^{k})}{8\pi G_{\!D}}\ell
_{\!e\!f\!\!f}^{2n-2}
\frac{(2n\!-\!1)!!^{2}}{(2n)!}\Big(1-\frac{2\alpha }{\ell
_{\!e\!f\!\!f}^{2}} (D\!-\!2)(D\!-\!3)\!\Big)\!,  \label{E0EGBodd}
\end{equation}
that was first computed in \cite{Kofinas-Olea} using Kounterterms
regularization. The form of the boundary terms that makes possible
this result for EGB-AdS gravity shall be shown to be universal below
because it also provides finite expressions for the conserved
quantities of AAdS solutions in Lovelock gravity.

The existence of a vacuum energy does not modify the black hole
entropy because as the total energy ${\mathcal{E}}=M+E_{0}$ is
shifted by a constant with respect to the mass calculated with
background-dependent methods, the Euclidean action changes in a
consistent manner. As a consequence, the entropy of the system can
be written as
\begin{equation}
S=\frac{vol(\Sigma
_{D-\!2}^{k})\,r_{+}^{D-2}}{4G_{\!D}}\Big[1+\frac{2k\alpha
(D\!-\!2)(D\!-\!3)}{r_{+}^{2}}\Big],  \label{SEGB}
\end{equation}
in both odd and even dimensions. This formula have been found by
several authors \cite{cai,ross, padma}, where some of the conserved
quantities, including the entropy function have been computed
assuming that they satisfy the First Law of black hole
thermodynamics. The same result can be derived from the regularized
Euclidean action as the free energy, obtained as difference between
the Euclidean bulk action evaluated for a EGB-AdS black hole and AdS
vacuum \cite{cho-neupane,neupane1,dut-gop} (for a similar
background-substraction computation in string generated gravity with
quadratic curvature couplings, see \cite{neupane2}).

\textbf{Dimensionally Continued Gravity.} If one considers that the
equation of motion for Lovelock gravity (\ref {EOMlovelock})
posseses $m$ different vacuum (constant curvature) solutions, this
means that $\alpha _{p}=0$ for $p>m$, while $\alpha _{m}\neq 0$ for
$ 1\leq m\leq \left[ \frac{D-1}{2}\right] $. Then,
eq.(\ref{EOMlovelock}) can also be written in the form
\begin{equation}
E_{\mu }^{\nu }=\,\delta _{\left[ \mu \mu _{1}\cdots \mu
_{2m}\right] }^{ \left[ \nu \nu _{1}\cdots \nu _{2m}\right]
}\,\left( \hat{R}_{\nu _{1}\nu _{2}}^{\mu _{1}\mu _{2}}+\gamma
_{1}\delta _{\lbrack \nu _{1}\nu _{2}]}^{[\mu _{1}\mu _{2}]}\right)
\cdots \left( \hat{R}_{\nu _{2m-1}\nu _{2m}}^{\mu _{2m-1}\mu
_{2m}}+\gamma _{m}\delta _{\lbrack \nu _{2m-1}\nu _{2m}]}^{[\mu
_{2m-1}\mu _{2m}]}\right) =0,  \label{EOMalter}
\end{equation}
where
\begin{equation}
\alpha _{m-p}=\alpha
_{m}\frac{(D\!-\!2m\!+\!2p\!-\!1)!}{(D\!-\!2m\!-\!1)!}
\sum_{i_{1}<...<i_{p}=1}^{m}\gamma _{i_{1}}\ldots \gamma
_{i_{p}}\,\,\,\,\,\,\,\,\,\,\,,\,\,\,\,\,\,\,\,\,\,\,1\leq p\leq m.
\label{alpham-p}
\end{equation}
The relation (\ref{alpham-p}) defines an algebraic system of $m$
equations for $m$ unknows $\gamma _{1},...,\gamma _{m}$. In the
particular case where $ \gamma _{1}=...=\gamma _{m}=\frac{1}{\ell
_{\!e\!f\!\!f}^{2}}$, the above equation produces for the couplings
$\alpha _{p}$ the special values
\begin{equation}
\alpha _{p}=\frac{(D\!-\!2p\!-\!1)!}{(D\!-\!3)!\,m}\ell
_{\!e\!f\!\!f}^{2p-2}\Big(\!\!
\begin{array}{c}
m \\
p%
\end{array}
\!\!\Big)\,\,\,\,\,\,\,\,\,\,\,,\,\,\,\,\,\,\,\,\,\,\,0\leq p\leq m.
\label{alphapmLUV}
\end{equation}
In the conventions we have adopted in this paper ($\alpha
_{0}=-2\Lambda $), we find that the effective AdS radius is $\ell
_{\!e\!f\!\!f}^{2}=\ell ^{2}/m$, whereas equation (\ref{elleff})
becomes an identity.

The label $m$ takes the maximal value ($m=[\frac{D-1}{2}]$) for two
particular Lovelock theories that feature a symmetry enhancement
from Lorentz to AdS group: Chern-Simons-AdS (CS-AdS) and
Born-Infeld-AdS (BI-AdS) in odd and even dimensions, respectively.
Both theories posses a single cosmological constant and the maximal
number of curvatures for a given dimension. Static black hole
solutions for CS-AdS and BI-AdS theories were studied in
\cite{DCBH}.

CS-AdS gravity is obtained from a Chern-Simons density for the AdS
group in $ D\!=\!2n+1$, such that the corresponding coefficients
(\ref{alphapmLUV}) in eq.(\ref{lovelock}) are given by
\begin{equation}
\alpha _{p}^{(C\!S)}=\frac{(D\!-\!2p\!-\!1)!}{(D\!-\!3)!\,n}\ell
_{\!e\!f\!\!f}^{2p-2}\Big(\!\!
\begin{array}{c}
n \\
p
\end{array}
\!\!\Big)\,\,\,\,\,\,\,\,\,\,\,,\,\,\,\,\,\,\,\,\,\,\,0\leq p\leq n,
\label{alphapCS}
\end{equation}
which produce equations of motion where AdS vacuum is a zero of
$n-$th order. Topological static black holes were studied in
\cite{Cai-Soh}. The horizon radius is defined by the relation
(\ref{ko}) $\mu =\frac{1}{n\ell
_{\!e\!f\!\!f}^{2}}(r_{+}^{2}\!+\!k\ell _{\!e\!f\!\!f}^{2})^{n}$,
such that the formula (\ref{massodd}) gives
\begin{equation}
M^{(C\!S)}=\frac{(D\!-\!2)\,vol(\Sigma _{D-\!2}^{k})}{16\pi n
G_{\!D}\ell_{\!e\!f\!\!f}^{2}}(r_{+}^{2}\!+\!k\ell_{\!e\!f\!\!f}^{2})^{n}
, \label{massCS}
\end{equation}
whereas the vacuum energy (\ref{Ezero}) reduces to the form%
\begin{equation}
E_{0}^{(C\!S)}=-k^{n}\frac{(D\!-\!2)\,vol(\Sigma
_{D-\!2}^{k})}{16\pi nG_{\!D}}\ell _{\!e\!f\!\!f}^{2(n-1)}.
\label{E0CS}
\end{equation}
The last expression corresponds to the energy of global AdS
spacetime. In CS-AdS gravity, the AdS vacuum is separated from black
holes ($M>0$) by a mass gap of naked singularities with mass in the
interval $M=(E_{0},0)$, as in $(2+1)$ dimensions. Eq.(\ref{massCS})
is the standard result for the mass, found in Hamiltonian form in
\cite{DCBH,Cai-Soh}. The vacuum energy was obtained as a Noether
charge evaluated in AdS in the background-independent formulation
presented in \cite{Mora-Olea-Troncoso-Zanelli-CS}, using a boundary
term proportional to (\ref{B2ntensor}). It is remarkable that the
symmetry enhancement in this case, turns the Dirichlet counterterms
series exactly solvable from the divergent parts in the expansion of
the canonical variation of the action \cite{BOT}, and this allows an
explicit comparison between the Kounterterms procedure and the
Dirichlet regularization \cite{mis-oleDCG}.
\newline
As usual in Lovelock gravity, black hole entropy in CS-AdS theory
cannot be related to the horizon area, but just expressed from
eq.(\ref{entroodd}) in terms of $r_{+}$ as
\begin{equation}
S^{(C\!S)}=\frac{(D\!-\!2)\,vol(\Sigma
_{D-\!2}^{k})}{4G_{\!D}}\int_{0}^{r_{+}}\!\!\!\!dr\,(r^{2}\!+\!k\ell_{\!e\!f\!\!f}^{2})^{n-1}.
\label{SCS}
\end{equation}
The last result matches the one obtained using a mini-superspace
model in the canonical formalism \cite{DCBH,Cai-Soh}, and also the
prescription for the entropy as a given $(D\!-\!2)-$form integrated
at the horizon \cite{ross}.

For BI-AdS gravity in $D\!=\!2n$ dimensions, the couplings
(\ref{alphapmLUV}) become
\begin{equation}
\alpha _{p}^{(B\!I)}=\frac{(D\!-\!2p)!}{(D\!-\!2)!\,n}\ell
_{\!e\!f\!\!f}^{2p-2}\Big(\!\!
\begin{array}{c}
n \\
p
\end{array}
\!\!\Big)\,\,\,\,\,\,\,\,\,\,\,,\,\,\,\,\,\,\,\,\,\,\,0\leq p\leq
n\!-\!1. \label{alphapBI}
\end{equation}
BI-AdS gravity can naturally incorporate into the bulk piece of the
action (\ref{lovelock}) the (topological) Euler term
${\mathcal{E}}_{2n}=\epsilon _{A_{1}...A_{D}}
\hat{{\mathcal{R}}}^{A_{1}A_{2}}...\hat{{\mathcal{R}}}^{A_{2n-1}A_{2n}}$
with an appropriate weight factor $\alpha_{n}^{(B\!I)}$ arising from
(\ref{alphapBI}) for $p=n$. As the Euler term is locally equivalent
to the boundary term (\ref{b3}), the complete action
(\ref{lovelock}) is also written as
\begin{equation}
I_{D}^{(B\!I)}=\frac{1 }{16\pi n
G_{\!D}}\frac{\ell_{\!e\!f\!\!f}^{2n-2}}{(D\!-\!2)!\,2^{n}}\int_{\!M}\!\!d^{2n}x\sqrt{-\mathcal{G}}\,\delta
_{[ \nu _{1}\cdots \nu _{2n}]}^{[\mu _{1}\cdots \mu _{2n}]}\Big(\!
\hat{R}_{\mu _{1}\mu _{2}}^{\nu _{1}\nu
_{2}}\!+\!\frac{1}{\ell_{\!e\!f\!\!f} ^{2}}\delta _{[ \mu _{1}\mu
_{2}]}^{[\nu _{1}\nu _{2}]}\!\Big) \cdots \Big(\!\hat{R}_{\mu
_{2n-1}\mu _{2n}}^{\nu _{2n-1}\nu
_{2n}}\!+\!\frac{1}{\ell_{\!e\!f\!\!f} ^{2}}\delta _{[ \mu
_{2n-1}\mu _{2n}]}^{[\nu _{2n-1}\nu _{2n}]}\!\Big), \label{BIAdS2n}
\end{equation}
which is both invariant under AdS group and regularized by
construction. Again, the Kounterterms procedure provides a finite
answer for the mass from eq.(\ref {Mass2n})
\begin{equation}
M^{(B\!I)}=\frac{vol(\Sigma _{D-\!2}^{k})}{8\pi
G_{\!D}\ell_{\!e\!f\!\!f} ^{2}} r_{+}
(r_{+}^{2}\!+\!k\ell_{\!e\!f\!\!f} ^{2})^{n-1},
\end{equation}
that has been obtained in Hamiltonian way, but also in a
background-independent method using the regularizing effect given by
the inclusion of the Euler term \cite{even}.
\newline
The static black hole entropy in BI-AdS gravity is found by plugging
the coefficients (\ref{alphapBI}) into the formula (\ref{S_even})
\begin{equation}
S^{(B\!I)}=\frac{vol(\Sigma
_{D-\!2}^{k})}{4G_{\!D}}[(r_{+}^{2}\!+\!k\ell_{\!e\!f\!\!f}
^{2})^{n-1}\!-\!(k\ell_{\!e\!f\!\!f} ^{2})^{n-1}]. \label{SBIAdS}
\end{equation}
The issue of the entropy for BI-AdS black holes is more subtle than
the regularization of the Noether charges. If, instead, one uses the
Euler term $ {\mathcal{E}}_{2n}$ to render the Euclidean action
finite as in (\ref{BIAdS2n}), the entropy found will be shifted by
the opposite of the last term proportional to $k^{n-1}$ in
(\ref{SBIAdS}), which is related to the Euler characteristic
$\chi_{2n}$ of the manifold. For solutions with hyperbolic horizon
($k\!=\!-1$), that entropy could become negative for physically
reasonable black holes ($r_{+}\!<\!\ell_{\!e\!f\!\!f} $), as noticed
for Einstein-Hilbert in \cite{OleaJHEP}. However, in our approach,
the problem is circumvented by using the Chern form (\ref{b3}), what
provides the consistent regularization prescription and the correct
entropy in all cases of even-dimensional Lovelock gravity.

\textbf{Lovelock Unique Vacuum.} Extending the idea of a single
cosmological constant of Dimensionally Continued AdS Gravity, one
can adjust the coefficients of Lovelock series to attain a family of
inequivalent gravity theories that posses a unique AdS vacuum
\cite{BHScan}. The choice (\ref{alphapmLUV}) produces equations of
motion where global AdS (maximally symmetric) spacetime is a zero of
$m-$th order.
\newline
The mass of asymptotically AdS static black holes was computed using
Hamiltonian formalism and AdS as the natural background reference
for the energy. Here, we use the background-independent formulas
(\ref{massodd}) and (\ref{Mass2n}), to obtain the mass
\begin{equation}
M^{(LUV)}=\frac{(D\!-\!2)\,vol(\Sigma _{D-\!2}^{k})}{16\pi m
G_{\!D}\ell _{\!e\!f\!\!f}^{2}}r_{+}^{D-2m-1}(r_{+}^{2}\!+\!k\ell
_{\!e\!f\!\!f}^{2})^{m}. \label{MLUV}
\end{equation}
In odd dimensions, the vacuum energy can be calculated directly from
eq.(\ref{Ezero}), and it turns to be
\begin{equation}
E_{0}^{(LUV)}=(-k)^{n}\frac{\,vol(\Sigma _{D-\!2}^{k})}{16\pi
nG_{\!D}}\frac{(2n\!-\!1)!!^{2} }{(D\!-\!3)!}\ell
_{\!e\!f\!\!f}^{2(n-1)}\int_{0}^{1}\!\!du\,u^{D-2m-1}(u^{2}\!-\!1)^{m-1}.
\label{e0LUV}
\end{equation}
The corresponding entropy can be computed from (\ref{entroodd}),
(\ref{S_even}), once the Euclidean action has been regularized by
the addition of the Kounterterms series, and takes the explicit form
\begin{equation}
S^{(LUV)}=\frac{(D\!-\!2)vol(\Sigma
_{D-\!2}^{k})}{4G_{\!D}}\int_{0}^{r_{+}}\!\!dr\,r^{D-2m-1}\,(r^{2}\!+\!k\ell_{\!e\!f\!\!f}^{2})^{m-1}.
\label{SLUV}
\end{equation}
In \cite{ross}, the above expression was obtained from the direct
application of Wald's prescription \cite{wald} for Lovelock Unique
Vacuum gravity. The same results can be reproduced using
identities derived from the gravitational bulk Lagrangian in
\cite{padma}. Despite the fact that these approaches lead to the
correct formula for the entropy in all cases, they deal only with
local properties of the action at the horizon and it do not really
provide an answer to the problem of bulk action regularization for
the asymptotic region in AdS gravity.
\newline
It is worthwhile noticing that this set of theories is not free from
the inconsistencies produced by negative values of the entropy
(\ref{SLUV}) when the spatial section has negative curvature. In
that sense, Lovelock Unique Vacuum does not feature a more sensible
thermodynamic behavior than, e.g., Einstein-Gauss-Bonnet with AdS
asymptotics.
\newline
The existence of different values for the vacuum energy
(\ref{e0LUV}) for a given odd dimension suggests that the set of
gravity theories ranging between EH and CS should have a set of
inequivalent CFT duals. This is also clear from
the information coming from the Weyl anomaly. On the contrary to EH-AdS, in $%
(2n\!+\!1)$-dimensional CS-AdS gravity the holographic Weyl anomaly
is proportional only to the Euler term in 2n-dimensions (type A
anomaly) with no contributions from the Weyl tensor (type B anomaly)
\cite{BOT}. Then, odd-dimensional gravity theories with $1<m<n$
should posses a combination of both types of holographic anomaly. As
this information is usually extracted from the finite part of a
quasilocal stress tensor for AdS gravity, the present regularization
prescription for all Lovelock theories can be regarded as a step
ahead towards a general formula for the holographic anomaly in
Lovelock-AdS gravity.

\section{Conclusions}

In this paper we have provided the explicit form of the boundary
terms that regularize the conserved quantities for asymptotically
AdS solutions of Lovelock gravity. The prescription for the boundary
terms contains the extrinsic curvature and it only distinguishes
even from odd dimensions, independently of the particular model
under consideration. Just the weight factor of these terms needs to
be consistently tuned in order to have a well-posed variational
principle for AAdS spacetimes. At the same time, the finiteness of
the Euclidean action is achieved.

In all the known cases (Einstein-Gauss-Bonnet, Chern-Simons,
Born-Infeld, Lovelock Unique Vacuum) both conserved charges and
black hole thermodynamics agree with the standard results. Even if
the Noether charges assign a non-vanishing vacuum energy to AdS in
odd dimensions (which is unobservable in background-dependent
methods), the entropy expression is still the correct one, because
the Euclidean action appears shifted consistently.

In even dimensions the boundary prescription is given by the maximal
Chern form. This is the structure appearing in the Euler theorem as
the correction to the Euler characteristic of the manifold due to
the boundary. In odd dimensions, the regularizing terms are linked
to the existence of extensions of Chern-Simons densities called
transgression forms \cite{trans}.

\[
\]
\textbf{Acknowlegements} G.K. is supported in part by EU grants
MRTN-CT-2004-512194 and MIF1-CT-2005-021982. R.O. is supported by
INFN. We thank G. Barnich, G. Compere, S. Detournay, J. Edelstein,
D. Klemm, O. Mi\v{s}kovi\'{c}, P. Mora and K. Skenderis for useful
conversations.

\end{document}